# Superconducting MoN thin films prepared by DC reactive magnetron sputtering for nanowire single-photon detectors

**Lily Hallett, Ilya Charaev, Akshay Agarwal, Andrew Dane, Marco Colangelo, Di Zhu, and Karl K. Berggren**

Department of Electrical Engineering and Computer Science, Massachusetts Institute of Technology,
50 Vassar Street, Cambridge, MA 02139, USA

E-mail: charaev@mit.edu

## Abstract

We present a comprehensive study of molybdenum nitride (MoN) thin film deposition using direct current (DC) reactive magnetron sputtering. We have investigated the effect of various deposition conditions on the superconducting and electrical properties of the films. Furthermore, we have shown that meander-shaped single-photon detectors made from 5 nm MoN films have saturated quantum detection efficiency at the telecom wavelength of 1550 nm. Our results indicate that MoN may be a material of interest for practical applications of low-temperature superconductors, including single-photon detectors and transition-edge sensors.

## 1. Introduction

Superconducting nanowire single-photon detectors (SNSPDs) have recently been implemented in many applications, including quantum information science and space-to-ground communication [1]. NbN is one of the most widely used material for these devices, but other materials, including MoN, may offer high internal detection efficiencies (IDE) and other desirable properties [2].

MoN has received attention in the last several decades for its mechanical properties, such as extreme hardness [3], and for the prediction of the B1 phase to be a 'high temperature' superconductor with a critical temperature of ~29 K [4–6]. While a material with a $T_c$ of 29 K would be of interest for high $T_c$ detector applications, the B1 phase is unfortunately thermodynamically and mechanically unstable, and does not appear on the MoN equilibrium phase diagram [7,8]. Many groups have attempted to achieve a metastable B1 phase by various techniques, all resulting in relatively low critical temperatures for as-deposited films [9–15]. The discrepancy between the experimental and theoretical $T_c$ can be attributed to nitrogen vacancies and defects [16–20], which can cause distortions on both the nitrogen and metal sublattices [21]. While the theoretical model takes nitrogen vacancies into account, it does not consider possible distortions of the metal sublattice [4]. The B1 phase was not achieved in this work, nor is it likely that the synthesis of a high $T_c$ B1 phase is feasible by any of the non-equilibrium techniques discussed in the literature thus far [22]. We have, however, succeeded in producing polycrystalline superconducting films by reactive DC magnetron sputtering, which were of sufficient quality to fabricate SNSPDs.

We have studied the electrical and superconducting properties of sub-20 nm MoN thin films depending on various deposition conditions, including the substrate, nitrogen partial pressure, temperature, discharge current, and post-deposition annealing. To our knowledge, there are only a couple of other reports of superconducting MoN films of less than 20 nm [23,24], which are of special interest for detector applications. We show that meander-shaped SNSPDs based on 5 nm MoN films demonstrate saturated internal detection efficiency at telecom wavelength. MoN thin films have been reported as a material for SNSPDs once before [25], but saturated detection efficiency was not achieved at telecom wavelength, and no information was given about the sputtering of the films or their crystal structure. In addition to demonstrating saturated detection efficiency, we provide a detailed report of our sputtering process and discuss several unique properties of MoN thin films.

## 2. DC reactive magnetron sputtering of MoN thin films

### *2.1 Discharge characteristics*

The MoN films were prepared by DC reactive magnetron sputtering using the AJA International ATC Orion Sputtering System. The films were deposited from the 2-inch pure Mo target (99.95%) in an argon (Ar) and nitrogen ($N_2$) atmosphere onto substrates which were placed without any thermal glue on a heater. A Si substrate with a 300 nm thermal oxide $SiO_2$ layer was used for all depositions in which the varied parameter was not the substrate. This substrate will hereafter be referred to as $SiO_2$. The substrates were cut into squares with an area of 1 cm$^2$ from a larger wafer and cleaned prior to deposition. The surface of the Mo target was pre-cleaned in a pure Ar atmosphere at a pressure of 2.5 mTorr. Before deposition, the plasma was stabilized in a mixed Ar and $N_2$ atmosphere. Figure 1 shows current-voltage (IV) discharge curves plotted for different nitrogen flow rates in standard cubic centimeters per minute (sccm). The flow rate of argon was held constant at 30 sccm. For all flow rates of nitrogen, the voltage increased nearly monotonically with increasing current.

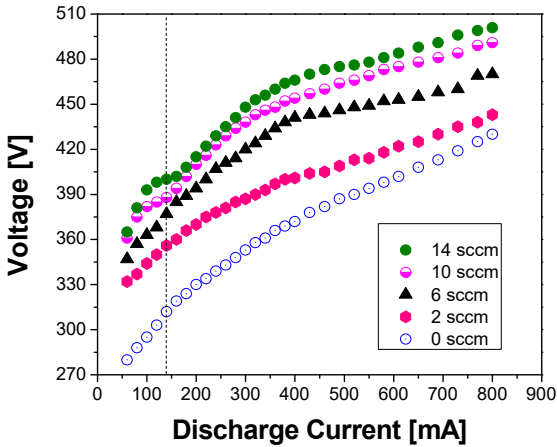

**Figure 1.** Discharge characteristics of the plasma in the sputtering chamber are shown for various nitrogen flow rates. For flow rates of 14 and 10 sccm, inflection points around 140 mA (marked by dashed line) were used to guide the choice of deposition parameters. Though there are small changes in slope at higher currents, films sputtered at high currents were not superconducting (See Figure S1b).

The IV curves relate to the relative rates of two processes: the chemical process in which metal-nitride forms on the Mo target and the physical process of sputtering material from the target to the substrate [26]. The chemical process of nitride formation is dependent on the partial pressure of nitrogen (controlled by nitrogen flow rate) and is unaffected by the discharge current, while the sputtering rate increases with increasing discharge current. It is crucial to select nitrogen flow rates and discharge currents such that a balance is struck between the rates of these competing processes.

For NbN, the optimal range of discharge currents for deposition at a given nitrogen partial pressure is in the region where the slope of the IV curve (the resistance) has a negative value [26]. In this region, the rates of the chemical and physical processes are equal. In our study of MoN, the slope of the IV curve is never truly negative, but it is close to zero just past 140 mA for $N_2 = 10$ sccm and $N_2 = 14$ sccm. We used these values of nitrogen flow rate and discharge current to guide the selection of deposition parameters. While the slope of the curves also changes slightly at higher currents, these changes are less pronounced. Furthermore, films deposited at high currents were not superconducting.

## 2.2 Variation of deposition parameters

We first performed a calibration of the deposition rate as a function of discharge current. Thicknesses of sputtered films were measured by x-ray reflectivity (Rigaku SmartLab) measurements. A linear increase in film thickness was observed, as higher currents lead to higher deposition rates. The relationship between the deposition rate and the discharge current is shown in the supplemental Figure S1a. Films sputtered at higher currents had lower $T_c$ despite being thicker, as demonstrated in Figure S1b. The sputtering rate at higher currents was likely faster than the rate of nitride formation, leading to pure Mo being sputtered onto the substrate, resulting in lower $T_c$.

Based on the deposition rate calibration and IV discharge characteristics, we sputtered a series of 18 nm thick MoN films at 140 mA. Multiple parameters were tuned during the deposition process to investigate the effects of these parameters on the critical temperature and electrical properties of the films. The parameters are listed in Table 1.

**Table 1.** A list of parameters that were varied during the deposition of 18 nm films.

| Parameter | Range of Variation |
|---|---|
| Growth Temperature [°C] | 25 - 840 |
| Argon Flow [sccm] | 27 - 39 |
| Nitrogen Flow [sccm] | 8 - 18 |
| Pressure [mTorr] | 2.5 - 4 |
| Substrate | GaN, MgO, Al$_2$O$_3$, SiN, SiO$_2$, Si |

The substrate temperature was varied to observe the effect of the growth temperature on the formation of stoichiometric MoN. Several 18 nm thick films were sputtered at 140 mA in the temperature range from 25 to 840°C. Figure 2a represents the dependence of the transition temperature on the growth temperature. The maximum $T_C$ was found at 500°C. In contrast to $T_C$, the sheet resistance $R_{sh}$ of MoN films decreased linearly with increasing temperature, as shown in Figure 2b. The $R_{sh}$ was reduced from 180 to 85 Ω as the temperature was raised from 25 to 840°C.

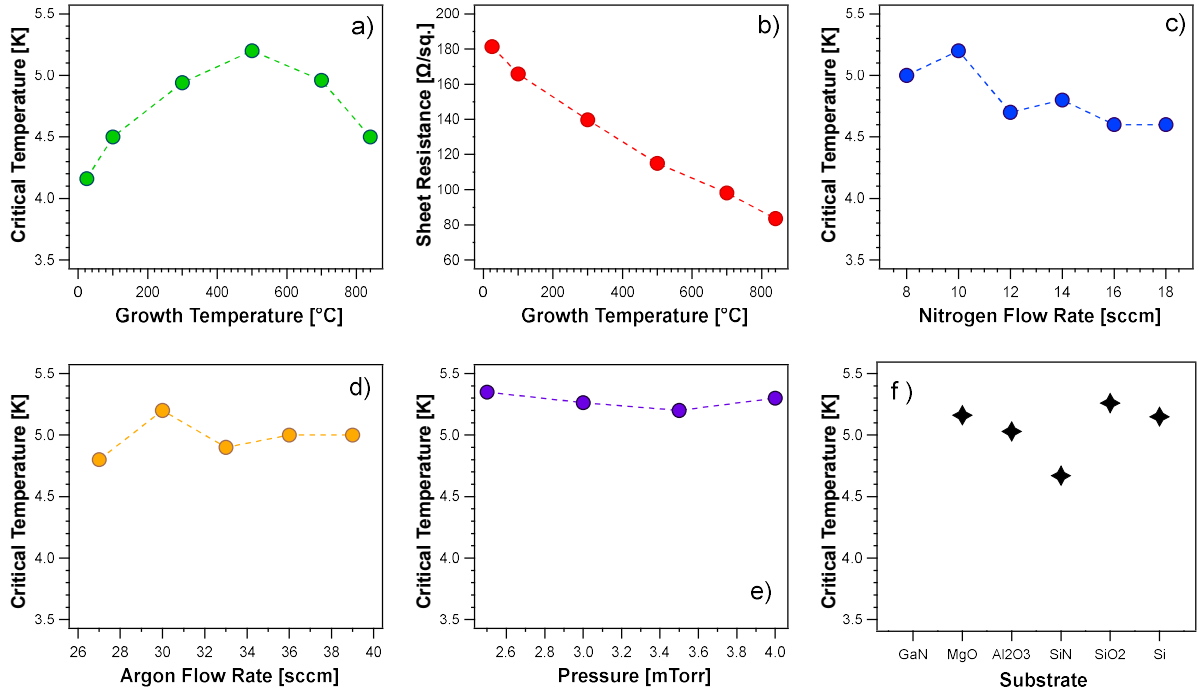

**Figure 2.** The standard deposition conditions for 18 nm films were: substrate = SiO$_2$; growth temperature = 500°C; pressure = 3.5 mTorr; argon flow F(Ar) = 30 sccm; nitrogen flow F(N$_2$) = 10 sccm, discharge current = 140 mA. One parameter was varied at a time while the others remained constant. **a)** Critical temperature vs. growth temperature; the $T_c$ reached a maximum at 500°C; **b)** Sheet resistance vs. growth temperature; the sheet resistance decreased linearly with increasing temperature; **c)** Critical temperature vs. nitrogen gas flow rate; **d)** Critical temperature vs. argon flow rate; **e)** Critical temperature vs. pressure; **f)** Critical temperature for various substrates; $T_c$ was insensitive to substrate with the exception of the film sputtered on GaN, which was not superconducitng.

Next, the flow rates of nitrogen and argon were varied to observe their effect on $T_c$. First, the nitrogen flow was varied from 8 to 18 sccm, while all other conditions were held constant (Figure 2c). The N$_2$ flow rate of 10 sccm resulted in the highest critical temperature. The flow rate of argon was varied from 27 to 39 sccm with all other standard parameters held constant. The highest $T_C$ occurred at an argon flow rate of 30 sccm (Figure 2d). A consistent trend between $T_C$ and argon flow rate was not observed. The deposition pressure was varied from 2.5 to 4 mTorr (Figure 2e). The variation of pressure had minimal effect on the critical temperature, but a pressure of 2.5 mTorr gave the highest $T_c$.

Depositions were performed on multiple substrates to investigate the effect of the seed layer on the electrical and superconducting properties of 18 nm MoN films (Figure 2f). The following substrates were used: gallium nitride (GaN), magnesium oxide (MgO), aluminum oxide (Al$_2$O$_3$), silicon nitride (SiN), silicon oxide (SiO$_2$), and silicon (Si). Surprisingly, the substrate used had minimal effect on the critical temperature and other properties, except for the film sputtered on GaN, which was not superconducting. MgO and SiO$_2$ showed slightly higher transition temperatures than the other substrates.

The variation of deposition parameters shown in Figure 2 was performed for 18 nm films to achieve critical temperatures within the measuring capability of our cryostat (>3.5 K). However, sub-10 nm films are preferable for SNSPD fabrication to attain high-sensitivity devices. In light of this, the last study performed before choosing deposition conditions for device fabrication was the variation of discharge current for 5 nm films. The discharge current was varied in a narrow range around 140 mA, as shown in Figure 3. A discharge current of 160 mA resulted in the highest $T_c$.

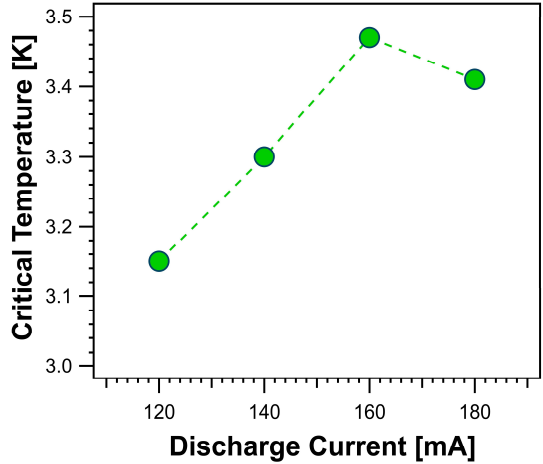

**Figure 3.** The discharge current was varied for a series of 5 nm films in a narrow range around 140 mA to find the optimum discharge current for thin films. A discharge current of 160 mA gave the highest critical temperature.

The transition temperatures of the 5 nm films were found to be lower than the measuring capability of the cryostat, but the beginnings of the transitions were observed, and the resistance vs. temperature curves were fitted using the theory of Aslamazov and Larkin [27]. These fits were used to calculate the critical temperatures of 5 nm films. Figure S2 shows an example of such a fitting.

Based on the results from Figures 2 and 3, we selected the deposition conditions for 5 nm films as follows: substrate = MgO; growth temperature = 500°C; pressure = 2.5 mTorr; argon flow F(Ar) = 30 sccm; nitrogen flow F(N$_2$) = 10 sccm, discharge current = 160 mA. MgO was chosen as the substrate because though it had a similar $T_c$ to SiO$_2$, MgO has a cubic structure and a lattice parameter of 0.42 nm, the same as B1 MoN. These conditions were used for the deposition of 5 nm films used in annealing experiments and device fabrication.

Motivated by the dramatic increase in the critical temperature of thick MoN films (> 20 nm) after annealing under a nitrogen atmosphere reported in [3,13,21,28], we performed annealing of MoN films at various temperatures and pressures. Annealing was performed by heating the sputtering chamber after depositions for several hours in a nitrogen atmosphere. Immediately after depositions were finished, temperature and pressure were gradually increased to anneal the films.

Table 2. Results from annealing experiments of 5 nm films.

| T, °C | t, hrs | P, mTorr | R, Ω/sq. | T$_C$, K |
|---|---|---|---|---|
| Not Annealed | -- | -- | 521 | 3.76 |
| 840 | 2 | 5 | 439 | 4.57 |
| 840 | 2 | 5 | 445 | 4.72 |
| 840 | 2 | 5 | 380 | 3.2 |
| 840 | 2 | 2.5 | 495 | 4 |
| 840 | 2 | 12 | 369 | 4.2 |
| 840 | 2 | 18 | 431 | 4.13 |
| 840 | 1 | 5 | 469 | 3.6 |
| 840 | 4 | 5 | 462 | 3.32 |
| 700 | 2 | 5 | 425 | 3.5 |

Annealing experiments performed on 5 nm films increased the critical temperature for certain conditions. The results of the annealing experiments for 5 nm films are presented in Table 2. The most successful annealing conditions of those we attempted were found to be a duration of 2 hours, a temperature of 840°C, and a pressure of 5 mTorr. The increase in $T_c$ provided by the annealing processes allowed us to fabricate SNSPDs with sufficiently high critical temperatures to perform optical measurements.

For 18 nm films, annealing processes had the effect of decreasing critical temperature, despite having sheet resistances that were approximately ten times lower than as-deposited films. This data can be found in the supplemental Table S2.

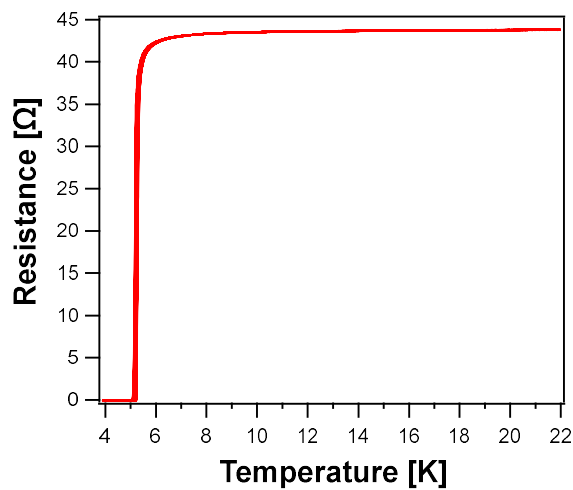

**Figure 4.** Resistance vs. temperature for an 18 nm film. The critical temperature of this film was 5.26 K. The plot above is a good representation of the transitions of all characterized films. All films had very sharp transitions.

It should be noted that the width of the superconducting transition (taken by subtracting the temperature at 90% of the resistance at 20 K by the temperature at 10% of the resistance at 20 K) was less than 0.3 K for nearly all characterized samples. A table containing the transition width ($\Delta T$) and deposition conditions of all samples for which $\Delta T$ was measured can be found in the supplementary materials in Table S1. Figure 4 shows one such transition for an 18 nm sample deposited at 3 mTorr, where all other deposition conditions were the standard conditions listed below Figure 2. The width of the transition was 0.2 K, which was close to the average width, 0.24 K, of all the characterized samples.

We have provided a detailed description of our sputtering experiments. These results were used to sputter 5 nm films with sufficiently high critical temperatures for device fabrication. Characterization of the films and a description of the device fabrication process will be discussed in subsequent sections.

*2.3 XRD and TEM analysis*

To study the morphological structure of MoN films, we used the X-Ray diffraction (XRD) analysis (Rigaku XRD SmartLab) with attenuator correction (high-resolution monochromator PB-Ge(220) × 2) on a thick 50 nm MoN film sputtered on an MgO substrate at the optimized conditions. The XRD data displayed in Figure 5a indicates the film is polycrystalline and shows a preferred orientation along the (111) plane; other weak peaks are (220) and (311) [29]. These peaks appear to be consistent with cubic $MoN_{0.46}$.

Additionally, we inspected 5 nm MoN films by TEM. TEM imaging was performed on a specially prepared MoN film sputtered onto a commercially available, 20 nm thick $Si_3N_4$ window. The TEM window was oxygen plasma cleaned ex-situ before being installed in the sputtering chamber. In-situ sputter cleaning was used for all TEM samples with the same settings as used for the other samples. Figure 5b shows TEM images of a 5 nm MoN film obtained by JEOL 2010F, which confirms the polycrystalline structure of our films. The small grain size indicates that MoN grew straight without any initial amorphous layers and has a uniform growth direction.

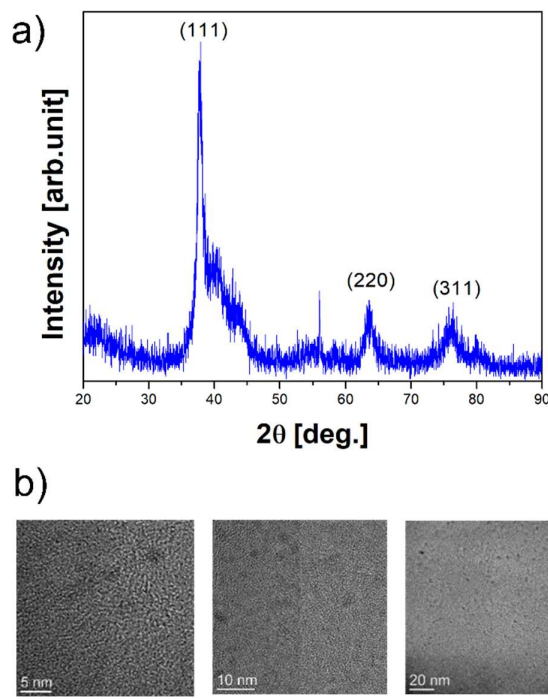

**Figure 5a)** X-ray diffraction from a ~50 nm thick MoN film deposited on an MgO substrate at a discharge current of 160 mA. The diffraction peaks are consistent with a cubic $MoN_{0.46}$ structure. **b)** TEM images of a 5 nm thick MoN film taken at different scales show the polycrystalline nature of the films and the small grain size.

## 3. SNSPDs

The superconducting SNSPDs were fabricated from 5 nm MoN films using electron-beam lithography with high-resolution positive electron-beam resist (ZEP 520 A). The resist was spin-coated onto the chip at 5000 rpm, which ensured a thickness of 335 nm. The resist was exposed to a 125 keV electron beam with an area dose density of 500 μC/cm$^2$. After exposure, the resist was developed by submerging the chip in O-xylene at 0°C for 90 s with subsequent rinsing in 2-propanol. The ZEP 520A pattern was then transferred to the MoN by reactive ion etching in $CF_4$ at 50W for 5 min.

The detectors were measured in a closed-cycle cryostat at 1.0 K. The samples were biased by a low-noise voltage source. The detector output was amplified using room-temperature LNA2500 amplifiers only. The amplified detector pulses were then either captured using a 6 GHz oscilloscope (Lecroy 760Zi, 40 G samples/s sampling rate) or a universal counter (Agilent 53132A). For optical measurements, a 1550 nm sub-picosecond fiber-coupled mode-locked laser (Calmar FPL-02CCF) was used.

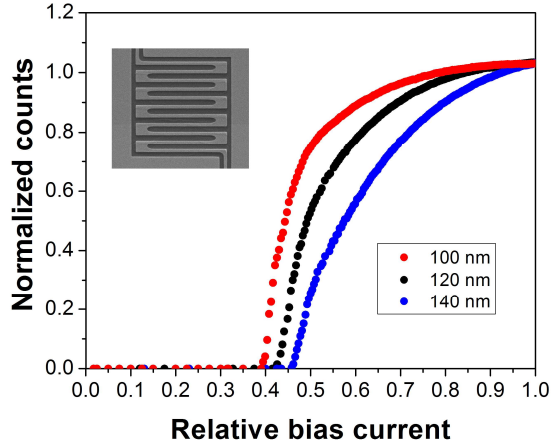

**Figure 6.** The count rate of MoN SNSPD detection events excluding dark counts as a function of the relative bias current. Data is plotted on a linear scale for widths of nanowires indicated in the legend. The plot shows that saturated detection efficiency was achieved. The inset shows an SEM image of the detector.

We tested SNSPDs with active areas of 4×4 μm$^2$ and widths ranging from 80 up to 120 nm. Figure 6 represents the optical response for nanowires with various widths indicated in the legend. Along with width-dependent single-photon sensitivity, 100 nm SNSPDs demonstrate the saturated internal detection efficiency at 1550 nm wavelength. The noise of the detectors (so-called dark count rate, *DCR*) was found to be below 10$^3$ cps. The timing jitter was measured in the range of 50 ps.

## 4. Discussion

In this work, we have performed a full sweep of the parameter space for the deposition of MoN films by DC reactive magnetron sputtering to investigate the effect of these parameters on the critical temperature and electrical properties of the films. High $T_c$ films were not achieved; however, we produced polycrystalline thin films with sufficiently high critical temperatures for use in SNSPDs. A detailed report of the results from the deposition experiments has been provided. A complete account of the results associated with the variation of deposition conditions is not present in the literature, to our knowledge. In the remainder of this section, we will discuss the results of the sputtering experiments, annealing experiments, and the SNSPD measurements.

In our experiments, the 5 nm films had significantly lower $T_C$ than the films of 18 nm. The decrease in $T_C$ with decreasing thickness of MoN has been discussed in other works [23]. Thin films of less than 10 nm are preferable for superconducting nanowire single-photon detectors. Despite the intrinsic proximity effect, our thin films have a critical temperature comparable with amorphous superconductors such as WSi and MoSi.

The substrate temperature influences the formation of stoichiometric and well-ordered MoN. Several works report 500°C to be the optimum deposition temperature, and this was also found to be true in our experiment. The critical temperature of the films increases with increasing growth temperature until reaching a maximum at 500°C. Sheet resistance decreases linearly with increasing temperature. The sheet resistance trend could reflect the nitrogen content in the films, with lower nitrogen content at higher temperatures resulting in more metallic and conductive films.

Higher nitrogen flow rates resulted in a degradation of $T_C$ in our films. In contrast to our results, in another report [30], increasing nitrogen content led to more stoichiometric films. The decrease in $T_C$ with increasing nitrogen flow rate in our experiments could be due to the fact that increasing nitrogen content can cause interstitial defects or a build-up of nitrogen at grain boundaries [19], while lower flow rates of nitrogen may have reduced interstitial defects and therefore reduced distortions on both sublattices.

The structure of the substrate will generally be imposed on the deposited film. Therefore, choosing a substrate with a similar structure to the desired phase of the film can reduce lattice mismatch and improve the structural quality of the films. Surprisingly, the substrate used had a minimal effect on the critical temperature and other parameters, though MgO and $SiO_2$ showed slightly higher transition temperatures. The insensitivity of the critical temperature to the substrate could indicate an amorphous film; however, XRD measurements and TEM imaging confirmed the polycrystalline nature of our MoN films. Furthermore, sputter-deposited films of MoN have never been reported to be amorphous. The insensitivity of MoN films to the substrate might be attractive for applications where the deposition of a superconductor on an unfavorable surface, along with stable electrical and superconducting properties, is required.

Another remarkable fact is the narrowness of the transition of the superconducting MoN films independent of stoichiometry, thickness, or seed layer. This feature could be of interest for transition-edge based detectors.

Several papers noted a dramatic increase in critical temperature after annealing under a nitrogen atmosphere [3,13,21,28]. It has been suggested that annealing could release interstitial nitrogen atoms bound at molybdenum sites, increasing order and critical temperature [21]. We attempted annealing experiments with the motivation to reduce possible disorder in our films. Increases in critical temperature by annealing have also been attributed to a phase conversion to δ-MoN at high pressure [28]. The δ-MoN phase is thermodynamically stable with a $T_C$ in the range of 13-15 K. The annealing experiments carried out in our sputtering system were performed at much lower pressures and likely resulted in nitrogen loss from the films [13] and no conversion to δ-MoN. In our experiments, performing heat treatments directly in the chamber at relatively low pressures after deposition eliminated the need for additional equipment such as a pressurized annealing furnace. While annealing 18 nm films under the conditions used in our study decreased critical temperature, 5 nm films showed an increase in $T_c$ after annealing, which allowed us to fabricate detectors and perform optical measurements.

Although a dramatic increase in critical temperature was not achieved by tuning deposition parameters, the single-photon detectors based on 5 nm MoN films demonstrated saturated detection efficiency at 1550 nm wavelength along with typical noise and timing characteristics for SNSPDs. MoN can be used as an alternative material for making single-photon devices for applications where low-temperature, high-performance detectors are required.

## 5. Conclusions

Parameters for the deposition of MoN films by DC reactive magnetron sputtering were systematically varied for a modern sputtering system. Results achieved for each set of conditions were clearly presented. Many previous reports of MoN thin films do not rigorously report their deposition conditions and results associated with changes in these conditions. We demonstrate the saturated internal detection efficiency along with typical noise and timing jitter characteristics for MoN SNSPDs.

Our results show that MoN could be a material of interest for practical applications of low-temperature superconductors. The saturated detection efficiency shown in this work indicates that MoN is a promising material for use in SNSPDs. The narrow width of the superconducting transition in thin MoN films might be a useful feature for sensors or transition-edge detectors that require this type of short transition width. Our films were insensitive to the choice of substrate, which could be an advantage when the deposition of a superconductor onto an unfavorable surface is required. This work highlights some of the unique properties of MoN, which may make it a useful alternative to more commonly used materials, like NbN, in low-temperature applications.

**Acknowledgments**

The authors would like to thank J. Daley and M. Mondol of the MIT Nanostructures lab for the technical support related to electron-beam fabrication. We would also like to thank Dr. C. Settens from the Center for Materials Science and Engineering X-ray Facility for his assistance and advice on all matters related to x-ray measurements, and Brenden Butters, Yujia Yang, Phillip Keathley and Mina Bionta for assistance in editing the final manuscript. We would also like to acknowledge the MIT Materials Research Laboratory Summer Internship Program. This research was sponsored by the U.S. Army Research Office (ARO) and was accomplished under the Cooperative Agreement No. W911NF-16–2-0192.

**References**

[1] C. M. Natarajan, M. G. Tanner, and R. H. Hadfield, "Superconducting nanowire single-photon detectors: Physics and applications," *Supercond. Sci. Technol.*, vol. 25, no. 6, 2012.

[2] W. Zhang *et al.*, "Saturating Intrinsic Detection Efficiency of Superconducting Nanowire Single-Photon Detectors via Defect Engineering," *Phys. Rev. Appl.*, vol. 12, no. 4, pp. 1–16, 2019.

[3] S. Wang *et al.*, "The Hardest Superconducting Metal Nitride," *Sci. Rep.*, vol. 5, pp. 1–8, 2015.

[4] D.A. Papaconstantopoulous, W.E. Pickett, B.M. Klein, and L.L. Boyer, "Nitride offers 30 Transition?," *Nat. Commun.*, vol. 308, no. 5, pp. 494–495, 1984.

[5] Z. You-xiang and H. Shou-an, "B1-Type MoN, A Possible High Tc Superconductor," *Solid State Commun.*, vol. 45, no. 3, pp. 281–283, 1983.

[6] W.E. Pickett, B.M. Klein, and D. A. Papaconstantopoulos, "Theoretical Prediction of MoN as a High Tc Superconductor," *Phys. 107B*, pp. 667–668, 1981.

[7] K. Balasubramanian, L. Huang, and D. Gall, "Phase stability and mechanical properties of Mo1-xNx with $0 \leq \times \leq 1$," *J. Appl. Phys.*, vol. 122, no. 19, pp. 0–12, 2017.

[8] H. Jehn and P. Ettmayer, "The Molybdenum Nitride Phase Diagram," *J. Less-Common Met.*, vol. 58, pp. 85–89, 1978.

[9] K. Saito and Y. Asada, "Superconductivity and structural changes of nitrogen-ion implanted Mo thin films," *J. Phys. F Met. Phys.*, vol. 17, no. 11, pp. 2273–2283, 1987.

[10] Y. H. Shi, B. R. Zhao, Y. Y. Zhao, L. Li, and J. R. Liu, "Superconducting and normal-state properties of MoNx thin films," *Phys. Rev. B*, vol. 38, no. 7, pp. 4488–4491, 1988.

[11] G. Linker, R. Smithey, and O. Meyer, "Superconductivity in MoN films with NaCl structure," *J. Phys. F Met. Phys.*, vol. 14, pp. L115-L119, 1984.

[12] H. Yamamoto, T. Miki, and M. Tanaka, "Preparation of Superconductive MoNx Films by Reactive Sputtering," *Adv. Cryog. Eng. Mater.*, pp. 671-672, 1986.

[13] N. Savvides, "High Tc superconducting B1 phase MoN films prepared by low-energy ion-assisted deposition," *J. Appl. Phys.*, vol. 62, no. 2, pp. 600–610, 1987.

[14] K. Inumaru, K. Baba, and S. Yamanaka, "Structural distortion and suppression of superconductivity in stoichiometric B1-MoN epitaxial thin films," *Phys. Rev. B - Condens. Matter Mater. Phys.*, vol. 73, no. 5, pp. 1–4, 2006.

[15] A. Bezinge, K. Yvon, J. Muller, W. Lengauer, P.Ettmayer, "High-Pressure High-Temperature Experiments on Delta-MoN," *Solid State Commun.*, vol. 63, no. 2, pp. 141–145, 1987.

[16] H. Ihara, Y. Kimura, K. Senzaki, H. Kezuka, and M. Hirabayashi, "Electronic structures of B1 MoN, fcc Mo2N, and hexagonal MoN," *Phys. Rev. B*, vol. 31, no. 5, pp. 3177–3178, 1985.

[17] L. Toth, "Superconducting Properites," in *Tranistion Metal Nitrides and Carbides*, J. Margrave, Ed. New York: Academic Press, 1971, pp. 217–222.

[18] H. Ihara, K. Senzaki, Y. Kimura, M. Hirabayashi, and N. Terada, "High Tc MoN Synthesis," *Adv. Cryog. Eng. Mater.* vol. 53, no. 9, pp. 603–616, 1986.

[19] L. Stöber, J. P. Konrath, V. Haberl, F. Patocka, M. Schneider, and U. Schmid, "Nitrogen incorporation in sputter deposited molybdenum nitride thin films," *J. Vac. Sci. Technol. A*, vol. 34, no. 2, pp.1-8, 2016.

[20] F. F. Klimashin, N. Koutná, H. Euchner, D. Holec, and P. H. Mayrhofer, "The impact of nitrogen content and vacancies on structure and mechanical properties of Mo-N thin films," *J. Appl. Phys.*, vol. 120, no. 18, 2016.

[21] G. Linker, H. Schmidt, C. Politis, R. Smithey, and P. Ziemann, "Magnetic susceptibility and defect structure of B1 phase MoN sputtered films," *J. Phys. F Met. Phys.*, vol. 16, no. 12, pp. 2167–2175, 1986.

[22] I. Jauberteau *et al.*, "Molybdenum nitride films: Crystal structures, synthesis, mechanical, electrical and some other properties," *Coatings*, vol. 5, no. 4, pp. 656–687, 2015.

[23] N. Haberkorn *et al.*, "Thickness dependence of the superconducting properties of γ- Mo2N thin films on Si (001) grown by DC sputtering at room temperature," *Mater. Chem. Phys.*, vol. 204, pp. 48–57, 2018.

[24] T. Tsuneoka, K. Makise, S. Maeda, B. Shinozaki, and F. Ichikawa, "Localization and pair breaking parameter in superconducting molybdenum nitride thin films," *J. Phys. Condens. Matter*, vol. 29, no. 1, 2017.

[25] Y. Korneeva *et al.*, "Comparison of Hot Spot Formation in NbN and MoN Thin Superconducting Films after Photon Absorption," *IEEE Trans. Appl. Supercond.*, vol. 27, no. 4, pp. 20–23, 2017.

[26] D. Henrich *et al.*, "Broadening of hot-spot response spectrum of superconducting NbN nanowire single-photon detector with reduced nitrogen content," *J. Appl. Phys.*, vol. 112, no. 7, 2012.

[27] L. G. Aslamazov and A. L. Larkin, "The Influence of Fluctuation Pairing of Electrons of the Conductivity of Normal Metal," *Phys. Lett.*, vol. 26A, no. 6, pp. 13–18, 1968.

[28] H. Ihara, M. Hirabayashi, K. Senzaki, Y. Kimura, and H. Kezuka, "Superconductivity of B1-MoN films annealed under high pressure," *Phys. Rev. B*, vol. 32, no. 3, pp. 1816–1817, 1985.

[29] C. L. Bull *et al.*, "Crystal structure and high-pressure properties of γ-Mo 2N determined by neutron powder diffraction and X-ray diffraction," *J. Solid State Chem.*, vol. 179, no. 6, pp. 1762–1767, 2006.

[30] T. Wang, G. Zhang, S. Ren, and B. Jiang, "Effect of nitrogen flow rate on structure and properties of MoNx coatings deposited by facing target sputtering," *J. Alloys Compd.*, vol. 701, pp. 1–8, 2017.

# Supplemental Material

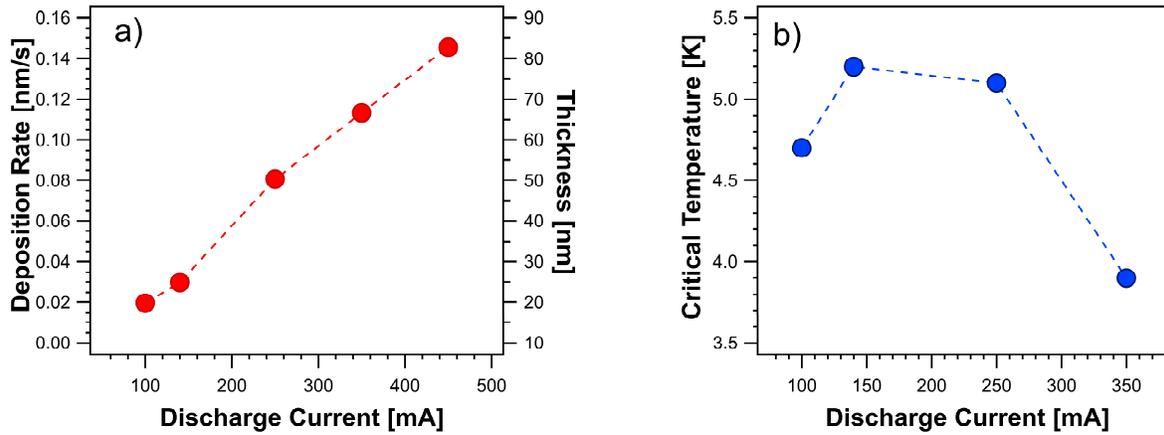

**Figure S1a)** Results of discharge current variation. All films were sputtered for 600 seconds. The deposition rate is shown on the left axis, and the thickness is shown on the right axis. The deposition rate is directly proportional to the thickness, which increases linearly with the increasing deposition rate. **b)** Critical temperature of films sputtered at different currents. Despite the fact that thicker films generally lead to higher $T_c$, films sputtered at higher currents had lower critical temperatures, and the film sputtered at 450 mA was not superconducting. Sputtering at higher currents likely resulted in more metallic films and lower $T_c$, as mentioned in the main text.

**Table S1.** The table below shows the deposition conditions for all samples for which the transition width (ΔT) was measured. The transition widths are listed from lowest to highest, but there is no clear relationship between the deposition conditions and the transition width. All transition widths are below 0.4, and the average ΔT of all characterized samples is 0.24 K.

| ΔT | Substrate | $T_c$ (K) | $F(N_2)$ (sccm) | Pressure (mTorr) | I (mA) | Rs (Ohm/sq) | RRR | Thickness, nm (XRR) |
|---|---|---|---|---|---|---|---|---|
| 0.09 | Si/SiO2 | 5.1 | 10 | 3.5 | 250 | 51.45 | 1.16 | 48 |
| 0.12 | Si/SiO2 | 5.3 | 10 | 3.5 | 140 | 90.58 | 2.00 | 22 |
| 0.12 | Si/SiO2 | 5 | 10 | 3.5 | 140 | 115.47 | 1.13 | 18 |
| 0.14 | Si/SiO2 | 3.9 | 10 | 3.5 | 350 | 27.27 | 2.83 | 68 |
| 0.18 | Si/SiO2 | 4.7 | 12 | 3.5 | 140 | 130.50 | 1.10 | 18 |
| 0.19 | MgO | 5.16 | 10 | 3.5 | 140 | 107.70 | 0.97 | 18 |
| 0.20 | Si/SiO2 | 5.26 | 10 | 3.5 | 140 | 104.43 | 0.93 | 18 |
| 0.21 | Si/SiO2 | 4.8 | 14 | 3.5 | 140 | 135.06 | 0.85 | 16 |
| 0.22 | SiN | 4.67 | 10 | 3.5 | 140 | 114.15 | 1.11 | 18 |
| 0.24 | Si/SiO2 | 4.9 | 10 | 3.5 | 140 | 118.18 | 0.74 | 18 |
| 0.26 | Si/SiO2 | 4.8 | 10 | 3.5 | 140 | 126.87 | 1.32 | 18 |
| 0.26 | Si/SiO2 | 5 | 10 | 3.5 | 140 | 111.99 | 0.34 | 18 |
| 0.26 | Si/SiO2 | 5.35 | 10 | 2.5 | 140 | 93.95 | 0.80 | 18 |
| 0.26 | Si/SIo2 | 4.6 | 18 | 3.5 | 140 | 146.33 | 0.95 | 18 |
| 0.28 | Si/SiO2 | 4.6 | 16 | 3.5 | 140 | 141.74 | 1.13 | 18 |
| 0.30 | Sapphire | 5.03 | 10 | 3.5 | 140 | 113.42 | 0.88 | 18 |
| 0.30 | SiO2 | 3.41 | 10 | 3.5 | 180 | 556.59 | 0.95 | 5 |
| 0.33 | Si/SiO2 | 5 | 8 | 3.5 | 140 | 112.41 | 1.21 | 18 |
| 0.35 | Si | 5.15 | 10 | 3.5 | 140 | 114.74 | 0.94 | 18 |
| 0.38 | Si/SiO2 | 4.7 | 10 | 3.5 | 100 | 182.25 | 1.06 | 12 |

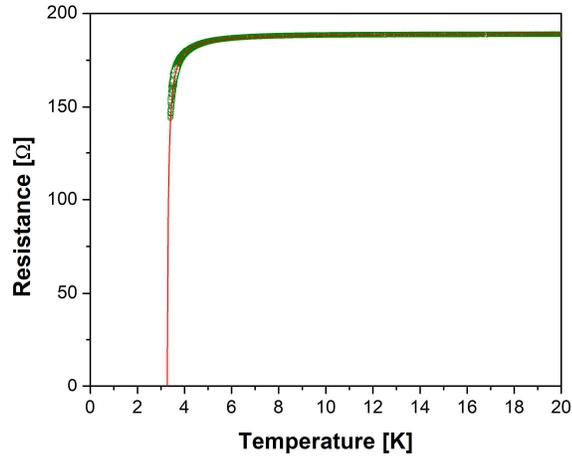

**Figure S2.** Resistance vs. temperature for a thin 5 nm film deposited at 160 mA, fitted with the theory of Aslamazov and Larkin [27]. Plots fitted in this way were used to calculate the critical temperatures of thin films whose full transitions were below the measuring capability of the cryostat (3.5 K).

**Table S2.** Parameters of annealing and properties of 18 nm MoN films after processing. Annealing processes on thicker films had the effect of decreasing critical temperature compared to films deposited under the same conditions that were not annealed.

| $T$, C° | $t$, hrs | $P$, mTorr | $R$, Ω/sq. | $T_C$, K | RRR |
|---|---|---|---|---|---|
| Not Annealed | -- | -- | 115.0 | 5.2 | 0.98 |
| 840 | 2 | 5 | 10.31 | 3.6 | 0.29 |
| 840 | 1 | 8 | 10.67 | <3.5 | 0.47 |
| 700 | 2 | 5 | 95.71 | 4.51 | 0.88 |